\documentclass[fleqn,usenatbib]{mnras}

\usepackage{newtxtext,newtxmath}
\usepackage[T1]{fontenc}

\DeclareRobustCommand{\VAN}[3]{#2}
\let\VANthebibliography\thebibliography
\def\thebibliography{\DeclareRobustCommand{\VAN}[3]{##3}\VANthebibliography}

\usepackage{graphicx}	% Including figure files
\usepackage{amsmath}	% Advanced maths commands
\usepackage{color}
\newenvironment{tightcenter}{%
	\setlength\topsep{0pt}
	\setlength\parskip{0pt}
	\begin{center}
}{%
 	\end{center}
}
\newcommand{\angstrom}{\textup{\AA}}

\title[Fundamental Planes for RLQ and RQQ]{The Fundamental Planes of Black Hole Activity for Radio-Loud and Radio-Quiet Quasars}

\author[Bariuan et al.]{Luis Gabriel C. Bariuan,$^{1,2}$\thanks{E-mail: lbariuan@mit.edu}
Bradford Snios,$^{2}$
Ma\l{}gosia Sobolewska,$^{2}$
\newauthor Aneta Siemiginowska,$^{2}$
Daniel A. Schwartz$^{2}$
 \\
$^{1}$Massachusetts Institute of Technology, Cambridge, MA 02139, USA\\
$^{2}$Center for Astrophysics $|$ Harvard \& Smithsonian, Cambridge, MA 02138, USA\\
}

\date{Accepted 2022 April 21. Received 2022 April 1;; in original form 2021 November 4}

\pubyear{2022}

\begin{document}
\label{firstpage}
\pagerange{\pageref{firstpage}--\pageref{lastpage}}
\maketitle

\begin{abstract}
We examine the fundamental plane of black hole activity for correlations with redshift and radio loudness in both radio-loud and radio-quiet quasar populations. Sources are compiled from archival data of both radio-loud and radio-quiet quasars over redshifts $0.1 < z < 5.0$ to produce a sample of 353 sources with known X-ray, radio, and black hole mass measurements. A fundamental plane of accretion activity is fit to a sample of radio-loud and radio-quiet quasars, and we find a dichotomy between radio-loud and radio-quiet sources. The set of best-fit equations that best describe the two samples are $\log{L_{R}} = (1.12 \pm 0.06) \log{L_{X}} - (0.20  \pm 0.07) \log{M} -(5.64 \pm 2.99)$ for our radio-loud sample and $\log{L_{R}} = (0.48 \pm 0.06) \log{L_{X}} + (0.50  \pm 0.08) \log{M} + (15.26 \pm 2.66)$ for our radio-quiet sample. Our results suggest that the average radio-quiet quasar emission is consistent with advection dominated accretion, while a combination of jet and disc emission dominates in radio-loud quasars. We additionally examine redshift trends amongst the radio-loud and radio-quiet samples, and we observe a redshift dependence for the fundamental plane of radio-loud quasars. Lastly, we utilize the fundamental plane as a black hole mass estimation method and determine it useful in studying systems where standard spectral modeling techniques are not viable.  
\end{abstract}


%%%%%%%%%%%%%%%%%%%%%%%%%%%%%%%%%%%%%%%%%%%%%%%%%%

%%%%%%%%%%%%%%%%% BODY OF PAPER %%%%%%%%%%%%%%%%%%

\section{Introduction} 
\label{sec:intro}
Black holes are exotic astrophysical objects with masses ranging from stellar mass up to $10^{10}\,M_{\odot}$, whose influence can be observed due to accretion processes that produce emissions across a broad range of different wavelengths. In particular, relativistic jets generated from the black hole will produce synchrotron radiation which is detectable at radio wavelengths \citep{Begelman1984}, while adiabatic compression within the inner accretion disc heats the surrounding material and produces X-ray emission \citep{Heinz1998,Stawarz2008}. Black hole luminosities can be as low as $10^{30}$\,--\,$10^{33}\rm\,erg\,s^{-1}$ in quiescent stellar mass black holes \citep[e.g.,][]{2003MNRAS.344...60G}, and as high as $10^{47}\rm\,erg\,s^{-1}$ in the most luminous quasars. By deriving correlations between the observables, black hole properties such accretion rate $\dot{M}$, disc-jet coupling, and jet models can be extrapolated \citep[e.g.,][]{2003A&A...400.1007C,2003MNRAS.344...60G}. Thus, multiwavelength observations of black holes are critical for studies of their unique physical properties.  

The observational study by \cite{Merloni2003} examined physical properties of low-redshift black holes ($z < 0.3$) over broad mass and luminosity ranges in X-ray and radio wavelengths for any evidence of correlations between those parameters. By studying $116$ black holes with known masses and luminosities, a ``fundamental plane of black hole accretion activity" characterized by $\log{L_{R}} = \xi_{RX} \log{L_{X}} + \xi_{RM} \log{M} +b_{R}$ was discovered.  The result demonstrated that the radio luminosity of a black hole has a direct dependence on its mass and X-ray luminosity, regardless of local environmental effects. Furthermore, the fundamental plane provides an empirical method of characterizing black hole behavior and properties that is scale invariant \citep{2003MNRAS.343L..59H,Merloni2003}. More recent works refined the fundamental plane parameters through the use of more precise radio and X-ray luminosities together with known masses obtained from direct dynamical measurements \citep{2006A&A...456..439K,Gultekin2009}. The tighter constraints led to a relation between the luminosities and mass that was consistent with previous fundamental plane measurements but with $33\%$ less scatter, further reinforcing the presence of a fundamental accretion process that is consistent across black holes in our nearby Universe. 
 
Despite the merit of previous results, the majority of fundamental plane studies prioritized black holes at low redshifts ($z<0.5$) since mass measurements and luminosities are well-constrained for those sources \citep[e.g.,][]{Merloni2003,Falcke2004,2006A&A...456..439K,Gultekin2009}. Increased cosmic densities within the early Universe may have driven periods of rapid formation due to increased merger rates and gas consumption, giving rise to different accretion physics than what is presently observed \citep{Volonteri2012}. In more recent timescales, the Universe has since undergone cosmological expansion which has also affected black hole formation and possibly accretion physics \citep{Merloni2003,2004MNRAS.353.1035M,2007MNRAS.380.1533M}. Thus, the validity of the fundamental plane is currently not well understood across a broad redshift range, resulting in a limited understanding of how black hole accretion evolved over large timescale. 

In addition to an unknown redshift dependence, several fundamental plane studies utilized a combination of radio-loud and radio-quiet sources in order to increase sample sizes and improve statistical errors \citep[e.g.,][]{Merloni2003,Falcke2004}. However, radio loudness has been shown to correlate with jet outflows and accretion activity in a black hole system \citep{2007ApJ...658..815S,2012ApJ...759...30B}. Thus, it is probable that the initial sample of black hole sources utilized in a fundamental plane study will significantly impact the best-fit results. Examination of a sample with a broad range of radio loudness is therefore important in understanding if a singular fundamental plane can be derived for all black holes, or if there is evidence that the fundamental plane evolves with radio loudness.

Motivated by these facts, we explored the observable properties of black holes across a broad redshift range in order to develop a fundamental plane of black hole activity. This derived relationship may then be compared with the relationship found amongst low-redshift sources to determine the presence, if any, of a redshift dependence in the fundamental plane. We additionally assemble a sample of radio-loud and radio-quiet quasars and examine differences between their respective fundamental plane models. This work complements recent studies of quasar structure and spectra performed across different wavelengths \citep[e.g.,][]{Just2007,Lusso2016,Vito2019b, 2020ApJ...899..127S}, by providing insights into the properties of black holes that formed at the early epochs of the Universe as well as the evolution of their  accretion properties.

The structure of the paper is as follows. In \S\,\ref{sec:dataselection}, we describe the sources we obtained from independent surveys. In \S\,\ref{sec:results}, we describe the fitting method for our fundamental plane analysis and present our best-fit results. \S\,\ref{sec:discussion} discusses observed trends in both redshift and radio loudness in our fundamental plane results as well as comment on utilizing the fundamental plane as method of estimating black hole mass. Our concluding remarks are discussed in \S\,\ref{sec:conclusion}.

For this paper, we adopted the cosmological parameters $H_{0} = 70\rm\,km\,s^{-1}\,Mpc^{-1}$, $\Omega_{\Lambda} =0.7$, and $\Omega_{M} = 0.3$ \citep{Hinshaw2013}. Estimated uncertainties are reported at a 1$\sigma$ confidence level, unless otherwise specified. 

\begin{table*}
    \caption{Properties of the Quasar Sample}
    \begin{tightcenter}
    \begin{tabular}{ccccccccccccc}
        \hline
        \hline
        Name & $z$ & log($R)$ & log($L_{R}$) & $\Delta$log($L_{R}$) &  Ref & log($L_X$) & $\Delta$log($L_X$) & Ref & log($M$) & $\Delta$log($M$)& Ref &log($L_X$/$L_{\rm Edd}$) \\
        \hline
        150436$-$024404&0.200&1.024&39.169&0.065&5&42.16&0.09&4&7.44&0.86&6&$-3.39$\\
        140051$+$025905&0.256&1.023&39.407&0.065&5&42.12&0.09&4&8.37&0.42&6&$-4.36$\\
        162901$+$400759&0.272&1.430&40.777&0.065&4&44.44&0.09&4&7.36&0.23&6&$-1.03$\\
        172255$+$320307&0.275&1.031&39.484&0.065&5&43.77&0.09&4&6.75&1.48&6&$-1.09$\\
        103336$+$573106&0.295&0.981&39.538&0.065&5&42.30&0.09&4&8.82&0.30&6&$-4.63$\\
        \hline 
    \end{tabular}
    \label{table:sample}
    \end{tightcenter}
    \raggedright{The first five sources from our compiled catalog of quasars are shown. The complete table is available online in machine-readable format.}\\
\end{table*}
 
\begin{figure*}
    \begin{tightcenter}
    \includegraphics[width=0.328\textwidth]{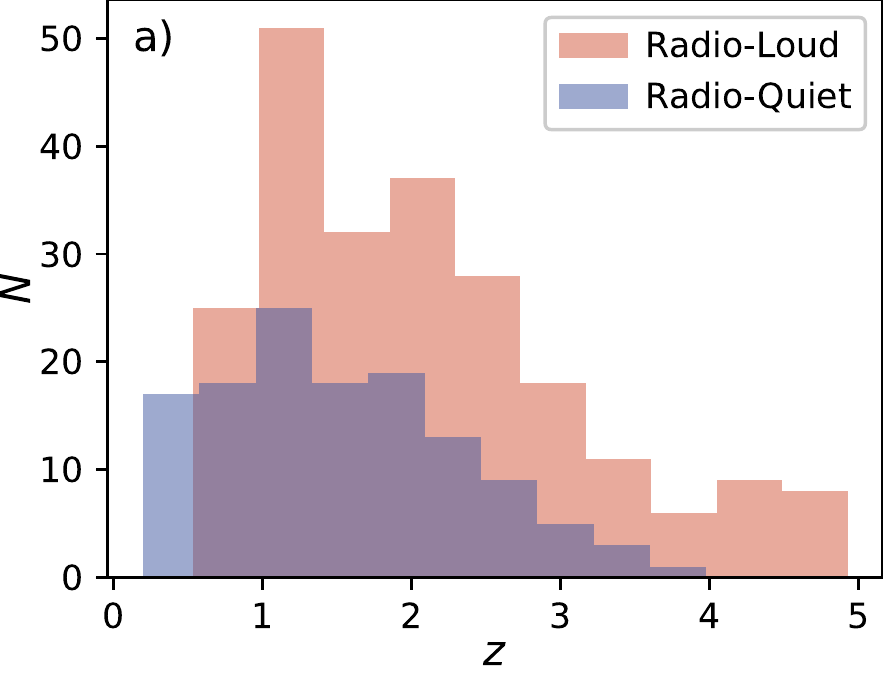}
    \hspace{0.05em}
    \includegraphics[width=0.328\textwidth]{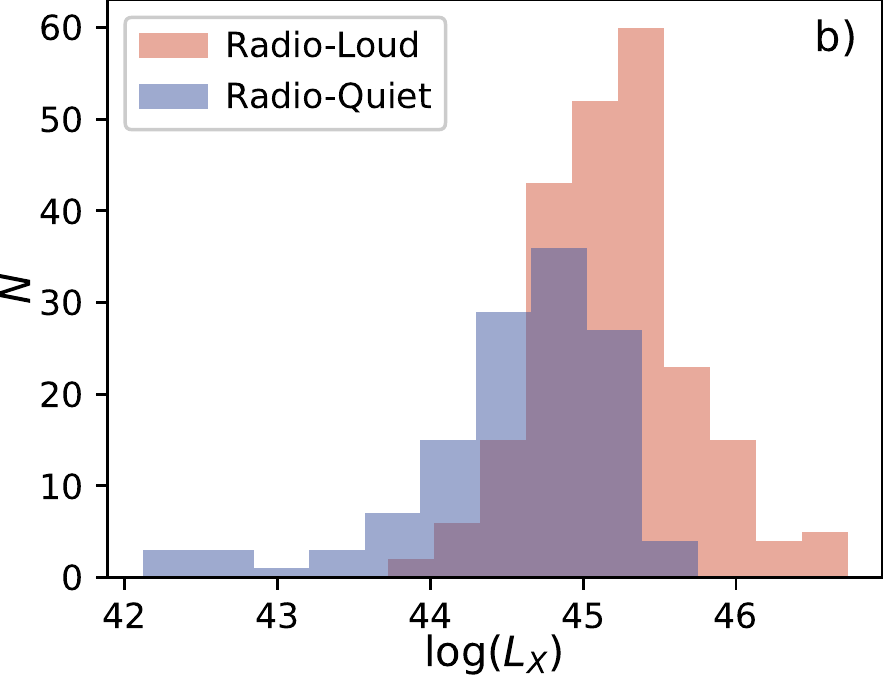}
    \hspace{0.05em}
    \includegraphics[width=0.328\textwidth]{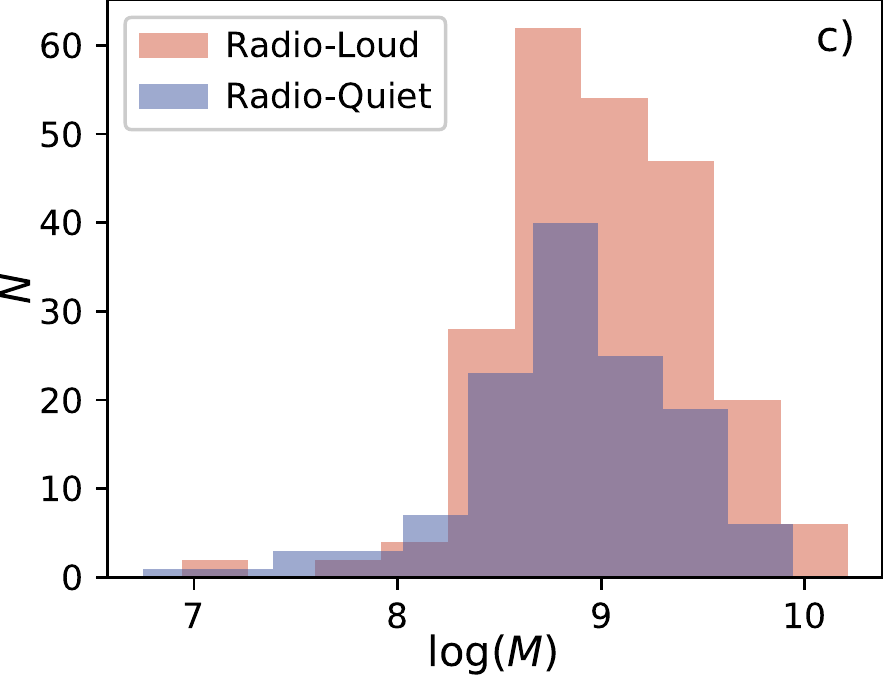}\\
    \vspace{1em}
    \includegraphics[width=0.328\textwidth]{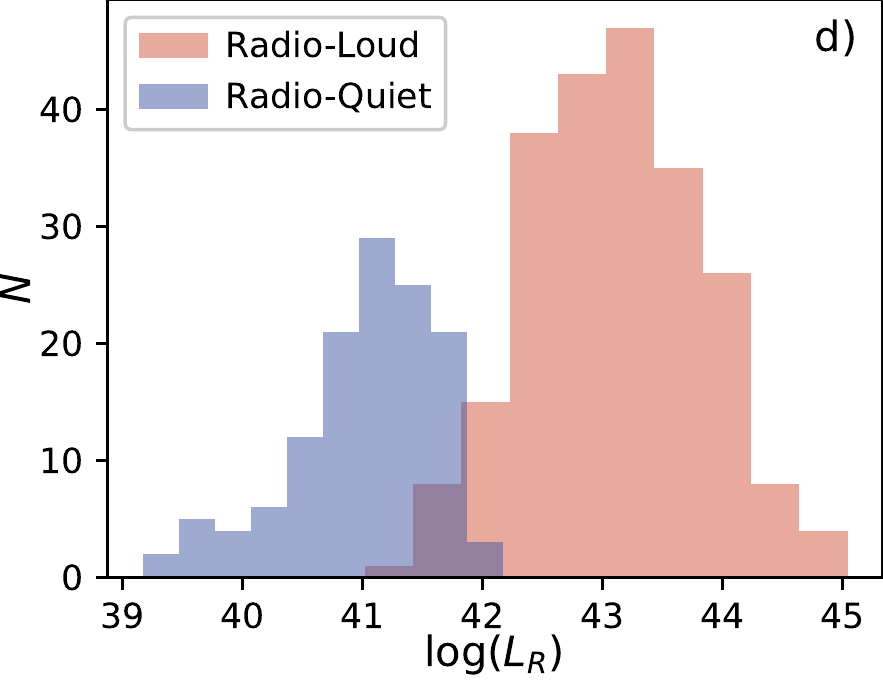}
    \hspace{0.05em}
    \includegraphics[width=0.328\textwidth]{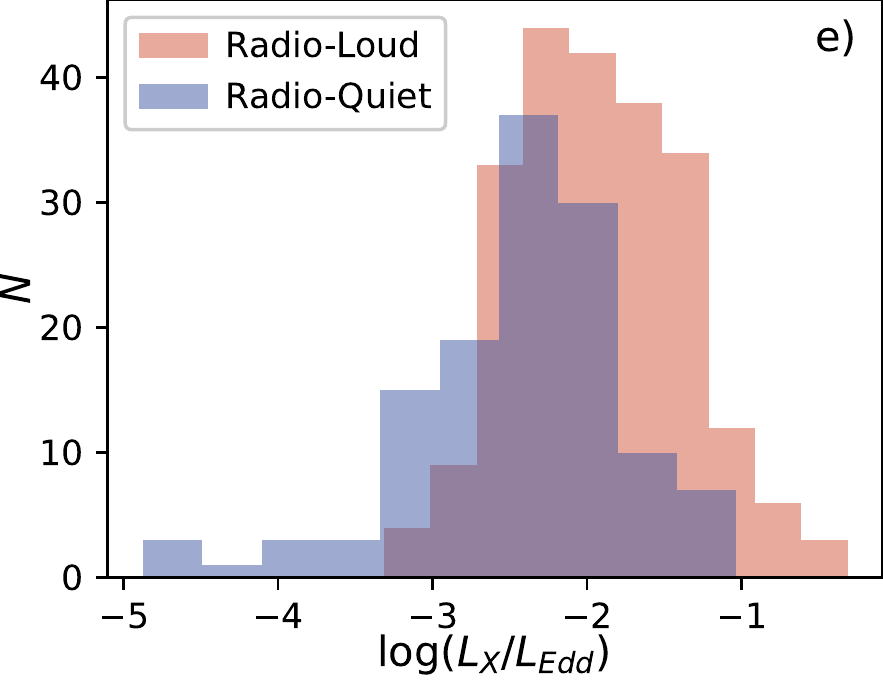}
    \hspace{0.05em}
    \includegraphics[width=0.328\textwidth]{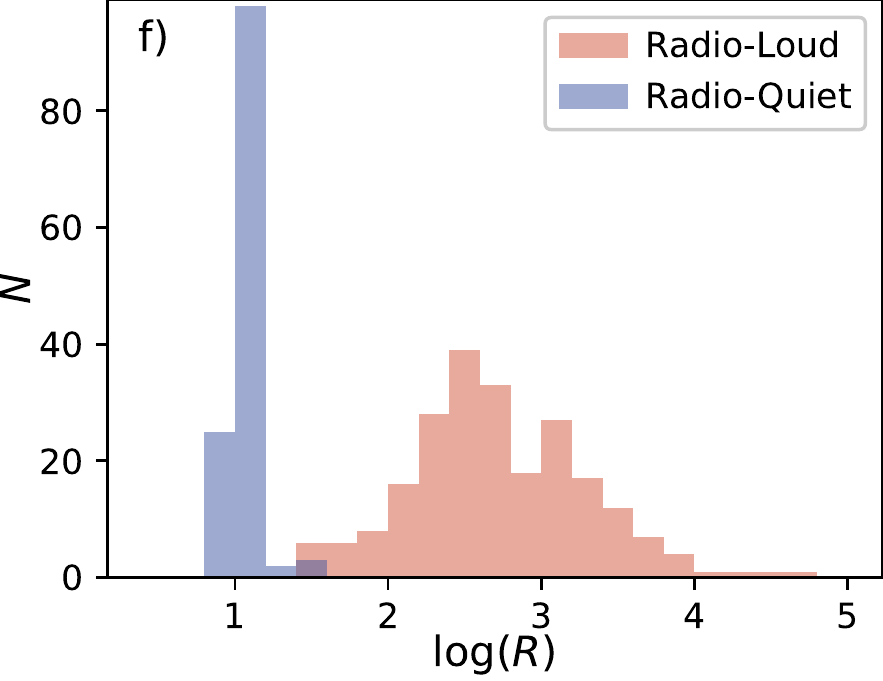}
    \end{tightcenter}
    \caption{Distributions of the radio-loud quasar (blue) and radio-quiet quasar (orange) samples used in our study of the fundamental plane for black hole activity. The histograms illustrate distributions across several different physical properties, including: (a) redshift $z$, (b) 2--10\,keV rest-frame X-ray luminosity log($L_{X}$), (c) spectroscopically measured black hole mass log($M$), (d) 5\,GHz rest-frame radio luminosity log($L_R$), (e) the ratio of X-ray luminosity to Eddington luminosity log($L_{x}/L_{\text{Edd}}$), and (f) radio loudness log($R$).
    }
    \label{fig:histogramsources}
\end{figure*}

%%%%%%%%%%%%%%%%%%%%%%%%%%%%
\section{Quasar Sample}
\label{sec:dataselection}

Previous investigations of quasars have indicated that radio loudness is inversely dependent to the accretion rate of the black hole system \citep{2007ApJ...658..815S,2012ApJ...759...30B}. It is therefore plausible that the derived fundamental plane of black hole activity for a sample will be impacted by the radio loudness distribution of its sources. We therefore require samples of both radio-loud and radio-quiet quasars with comparable physical properties in order to investigate dependencies on radio loudness in our fundamental plane analysis. 

For our study, we define radio-loudness according to the criterion described by \cite{1989AJ.....98.1195K} where radio-loud objects are defined as any source with a radio loudness\footnote{Radio loudness is defined as $R = f_{\rm5\,GHz}/f_{\rm4400\,\angstrom}$, where $f_{\rm5\,GHz}$ and $f_{\rm4400\,\angstrom}$ correspond to the rest-frame flux density at 5\,GHz and 4400\,\angstrom, respectively.} parameter $R>30$. Furthermore, we require that each sample must have a comparable sample size to previous works on the fundamental plane \citep[$>$\,100 sources,][]{Merloni2003,Falcke2004} to ensure comparable statistical precision. We therefore searched the literature for archival datasets that would supply us with a sufficient number of quasars spanning our desired physical parameter space.

%%%%%%%%%%%%%%%%%%%%%%%%%%%%%%%%%%%%%%
\subsection{Radio-Loud Quasars}
\label{subsec:radioloudquasars}

For our radio-loud sample of quasars, we gathered data across several different works that contained complete descriptions of radio and X-ray luminosities. Since a primary goal of our analysis is to also investigate redshift dependence in accretion activity, we included samples spanning a broad redshift range. Sources with X-ray and radio luminosity measurements were selected from the radio-loud quasar catalogs of \cite{2019MNRAS.482.2016Z}, \cite{2020ApJ...899..127S}, \cite{Snios2021}, and \cite{2021MNRAS.505.1954Z}, where the luminosity derivation methods utilized in those works are briefly summarized below. 

Radio fluxes were taken from the VLA-FIRST radio survey \citep{Becker1995}, or from independent VLA observations when FIRST data of the source was not available. A region was defined to isolate the AGN core in the radio observation, and the radio flux was measured from the region. The rest-frame 5\,GHz radio luminosity of each source was extrapolated from the flux measurements assuming a spectral index of $\alpha_{R} = -0.5$. Radio loudness parameters for the quasars from \cite{2019MNRAS.482.2016Z}, \cite{2020ApJ...899..127S}, and \cite{2021MNRAS.505.1954Z} were provided from the original sources of the data. For the \cite{Snios2021} sample, radio loudness values were derived by extrapolating the provided $f_{2500\,\text{\AA}}$ values to 4400\,\text{\AA} assuming a spectral index of $\alpha_{o}=-0.5$.

At the 5\arcsec\ resolution of FIRST, the survey resolved radio features up to a separation distance of 31\,kpc at $z=5$, the maximum redshift of our quasar sample. We note that radio emission in nearby sources studied in the context of the fundamental plane reflects activity of the nucleus and generally does not include the extended radio emission from large scale jets \citep[e.g.,][]{Merloni2003,Falcke2004}. Since a priority of our sample is to investigate high-redshift sources, we anticipate a possible bias in our sample due to the inclusion of extended radio emission in our measurements. We explore such possible bias in the discussion of our redshift analysis (\S~\ref{subsec:redshiftdepedence}).

X-ray luminosities were derived using observations from Chandra, XMM-Newton, and Swift. A region was defined around the AGN core, and an emission spectrum was extracted from the source. Extended X-ray features potentially present in the regions may be ignored as they generally provide a negligible contribution to the total quasar luminosity \cite[e.g.,][]{Marshall2018,Worrall2020}. Furthermore, the majority of quasars in our sample were observed in short exposures, which reduces the signal-to-noise of diffuse X-ray features and subsequently limits their impact on the extracted spectral profile of the AGN. The observed X-ray flux was measured from a spectral model of the 0.5--7.0\,keV energy band, and the rest-frame 2--10\,keV luminosity was extrapolated from the best-fit model. We specifically selected quasars with radio and X-ray flux measurements and omitted any sources with upper limit estimates. Our quasar catalog was cross-referenced with the SDSS-DR14 catalog \citep{2020ApJS..249...17R}, where sources with confirmed spectroscopic mass measurements via SDSS spectra were included in our fundamental plane analysis. 

Ultimately, our radio-loud quasar sample included 11 sources from \cite{2019MNRAS.482.2016Z}, 6 sources from \cite{2020ApJ...899..127S}, 12 sources from \cite{Snios2021}, and 196 sources from \cite{2021MNRAS.505.1954Z}. Radio luminosities for the sample ranged between $10^{40}$\,--\,$10^{46}$\,erg\,$\text{s}^{-1}$, while X-ray luminosities ranged between $10^{43}$\,--\,$10^{47}$\,erg\,$\text{s}^{-1}$. The masses of the sources ranged between $10^{6}$\,--\,$10^{11}\,M_{\odot}$ with an average uncertainty of $0.342\,M$. Consistent with \cite{Merloni2003}, we also calculated for each source the ratio of the X-ray luminosity to the Eddington luminosity, which may be used as a proxy for the Eddington ratio. We defined the ratio as $L_X/L_{\rm Edd}$ where $L_{\rm Edd}=1.3\times 10^{38} M/M_{\odot}$. The range of log($L_X/L_{\rm Edd}$) for our radio-loud sample is [$-3.31$,$-0.31$] with a mean of $-1.92$. In total, we obtained $225$ unique radio-loud quasars at $0.1 < z < 5.0$ with complete measurements over a broad range of observables, which is sufficient for our modeling analysis. We note that our sample includes quasars with known X-ray luminosity, radio luminosity, and mass. We found that inclusion of sources with luminosity upper limits added only 3 sources to our total, so we discarded upper limit sources from our sample for consistency with previous studies. Details on our radio-loud catalog are provided in Table~\ref{table:sample}, and Figure \ref{fig:histogramsources} shows the source distribution of our radio-loud sample across several different physical properties.

%%%%%%%%%%%%%%%%%%%%%%%%%%%%%%%%%%%%%%
\subsection{Radio-Quiet Quasars}
\label{subsec:radioquietquasars}

\begin{table*}
    \caption{Best-fit Parameters of the Fundamental Plane of Black Hole Activity Analysis}
    \label{table:fundplane}
    \begin{tabular}{lcrrrrrr}
        \hline
        \hline
         & Number of Sources & \multicolumn{1}{c}{$\xi_{X}$} & \multicolumn{1}{c}{$\xi_{M}$} & \multicolumn{1}{c}{$b$} & \multicolumn1c{$\sigma^2_{\rm res}$} & \multicolumn{1}{c}{$R_{\rm adj}^2$} & \multicolumn{1}{c}{$\sigma_{R}$}\\
        \hline
        Total Dataset & $353$ & $1.44 \pm 0.08$ & $0.16 \pm 0.10$ & $-23.85 \pm 3.49$ & $54.97$ & $0.5207$ & $0.85$\\
        \hline
        \textit{Radio Loudness Dependence}\\
        \hline
        Radio-Loud Quasars & $225$ & $1.12 \pm 0.06$ & $-0.20 \pm 0.07$ & $-5.64 \pm 2.99$ & $20.09$ & $0.4629$ & $0.62$\\
        Radio-Quiet Quasars & $128$ & $0.48 \pm 0.06$ & $0.50 \pm 0.08$ & $15.26 \pm 2.66$ & $17.77$ & $0.5866$ & $0.39$\\
        \hline
        \textit{Redshift-dependence (Radio-Loud Quasars)}\\
        \hline
        $z<1.5$ & $87$ & $1.32 \pm 0.15$ & $-0.06 \pm 0.13$ & $-16.20 \pm 6.59$ & $18.58$ & $0.3390$ & $0.55$\\
        $1.5<z<3$ & $94$ & $1.06 \pm 0.13$ & $-0.17 \pm 0.10$ & $-3.27 \pm 6.00$ & $21.66$ & $0.3564$ & $0.52$\\
        $z>3$ &$44$& $0.63 \pm 0.15$ & $-0.27 \pm 0.15$ & $17.55 \pm 6.64$ & $22.63$ & $0.2779$ & $0.48$\\
        \hline
        \textit{Redshift-dependence (Radio-Quiet Quasars)}\\
        \hline
        $z<1.5$ & $66$ & $0.38 \pm 0.08$ & $0.40 \pm 0.11$ & $20.20 \pm 3.32$ & $22.68$ & $0.5035$ & $0.42$\\
        $z>1.5$ & $62$ & $0.18 \pm 0.09$ & $0.22 \pm 0.08$ & $31.49 \pm 4.12$ & $11.17$ & $0.4460$ & $0.26$\\
        \hline
        \end{tabular}
        \\\raggedright{{\sc note}: The reported parameters for the fundamental plane best-fits correspond to the X-ray luminosity slope $\xi_{X}$, the black hole mass slope $\xi_{M}$, the y-intercept $b$, the best-fit residual variance $\sigma^2_{\rm res}$, the adjusted R-squared of the best-fit model $R_{\rm adj}^2$, and the scatter of the log$(L_R)$ parameter relative to the predicted value from the fundamental plane model $\sigma_{R}$.}
\end{table*}

Radio-quiet quasars comprise the majority of the known quasar population and are therefore commonly utilized in fundamental plane analyses \citep[e.g.,][]{Merloni2003,Falcke2004,2006A&A...456..439K,Gultekin2009,2012MNRAS.419..267P}. To obtain our radio-quiet quasar sample, we cross-referenced the radio-quiet quasar catalog from \cite{2020MNRAS.498.4033T} with the X-ray fluxes of \cite{2021MNRAS.505.1954Z}. Consistent with our sample selection method for the radio-loud catalog, we selected radio-quiet quasars with known radio and X-ray flux measurements. Radio and X-ray luminosities were provided from \cite{2020MNRAS.498.4033T} and \cite{2021MNRAS.505.1954Z}, where the derivation methods are the same as those described in \S~\ref{subsec:radioloudquasars}. Our quasar catalog was cross-referenced with the SDSS-DR14 catalog for spectroscopic mass measurements \citep{2020ApJS..249...17R}. The radio loudness parameters for our sample were provided from \cite{2020MNRAS.498.4033T}.

In total, we found $128$ radio-quiet quasars with complete measurements which we included in our fundamental plane analysis. Similar to the radio-loud sample, inclusion of radio-quiet quasars with luminosity upper limits added a negligible amount of sources to the catalog, 9 sources in total. Our radio-quiet sample of $128$ sources therefore omits sources with luminosity upper limits. We note that although radio-quiet quasars comprise the majority of the known quasar population, our radio-loud sample is larger than our radio-quiet sample because relatively few radio-quiet sources have radio fluxes above current detector sensitivity limits. The sources span a redshift range of $0.1 < z < 4.0$, X-ray luminosities that ranged between $10^{42}$\,--\,$10^{46}$\,erg\,$\text{s}^{-1}$, radio luminosities that ranged between $10^{39}$\,--\,$10^{43}$\,erg\,$\text{s}^{-1}$, and masses that ranged between $10^{6}$\,--\,$10^{10}\,M_{\odot}$ with an average uncertainty of $0.341\,M$. The ratio log($L_X/L_{\rm Edd}$) was also calculated for our radio-quiet sample, finding a range of [$-4.87$,$-1.03$] with a mean of $-2.43$. The radio-quiet sources possess similar physical parameters as our radio-loud sources, and both samples are comparable in size. This ensures that radio loudness is the primary difference between our two samples. Details on our radio-quiet catalog are provided in Table~\ref{table:sample}, Figure \ref{fig:histogramsources} shows the source distribution of our radio-quiet sample across different physical properties. 

%%%%%%%%%%%%%%%%%%%%%%%%%%%%%%%%%%%%%%
\section{Modeling the Fundamental Activity Plane}
\label{sec:results}
With our accumulated dataset from \S~\ref{sec:dataselection}, we constructed a fundamental plane of black hole activity. Uncertainties for sources from \citet{2020ApJ...899..127S,Snios2021} were provided in their work. Since sources found within \citet{2019MNRAS.482.2016Z,2020MNRAS.498.4033T,2021MNRAS.505.1954Z} did not include quoted errors, we adopted a $20 \%$ uncertainty for their X-ray luminosities. For all of our radio luminosities, we adopted a $15 \%$ uncertainty. We note that our adopted errors exceed those from similar studies \citep[e.g.,][]{Gultekin2009,2019ApJ...871...80G}, which reassures us that our error estimates are conservative. For the mass measurements obtained from \cite{2020ApJS..249...17R}, we utilized the quoted uncertainties. Due to our model fitting method (described below), we adopted conservative symmetric uncertainties for all measurements.

Consistent with \cite{Merloni2003}, we defined the fundamental plane of activity as
\begin{equation}
    \log{L_{R}} = \xi_{X}\log{L_{X}} + \xi_{M}\log{M}+b,
    \label{eqn:fundaplane}
\end{equation}
where $L_{R}$ is the rest-frame 5\,GHz luminosity, $L_{X}$ is the rest-frame 2--10\,keV luminosity, and $M$ is the black hole mass. The dataset was fit using the multivariate orthogonal distance regression (ODR) method to estimate the best-fit regression coefficients \citep{Boggs1990}. ODR is a fitting scheme that modifies the ordinary least squares by accounting for uncertainties within both the dependent and independent variables, and the data is minimized by the sum of squared perpendicular distances from the data points to the best-fit line weighted by the uncertainty. Given the presence of measurement uncertainties for our three input parameters, ODR provided a more accurate best-fit than the commonly used ordinary least square regression method. 

Despite the robustness of the ODR method, a single run is not guaranteed to produce well-constrained best-fit coefficients with our data set as the maximally strict convergence conditions of the sums of squares and relative changes in the estimated parameters yield fits that are not necessarily located at the global minimum. We therefore ran ODR with $10^4$ iterations, which produced a symmetric distribution of fitting coefficients. We used the mean of the resultant parameters as our best-fit coefficients, while their errors were calculated from the mean covariance matrices. We additionally implemented randomized initial conditions for each run to reduce input bias on the best-fit parameters. 

\begin{figure*}
    \centering
    \includegraphics[width=0.99\linewidth]{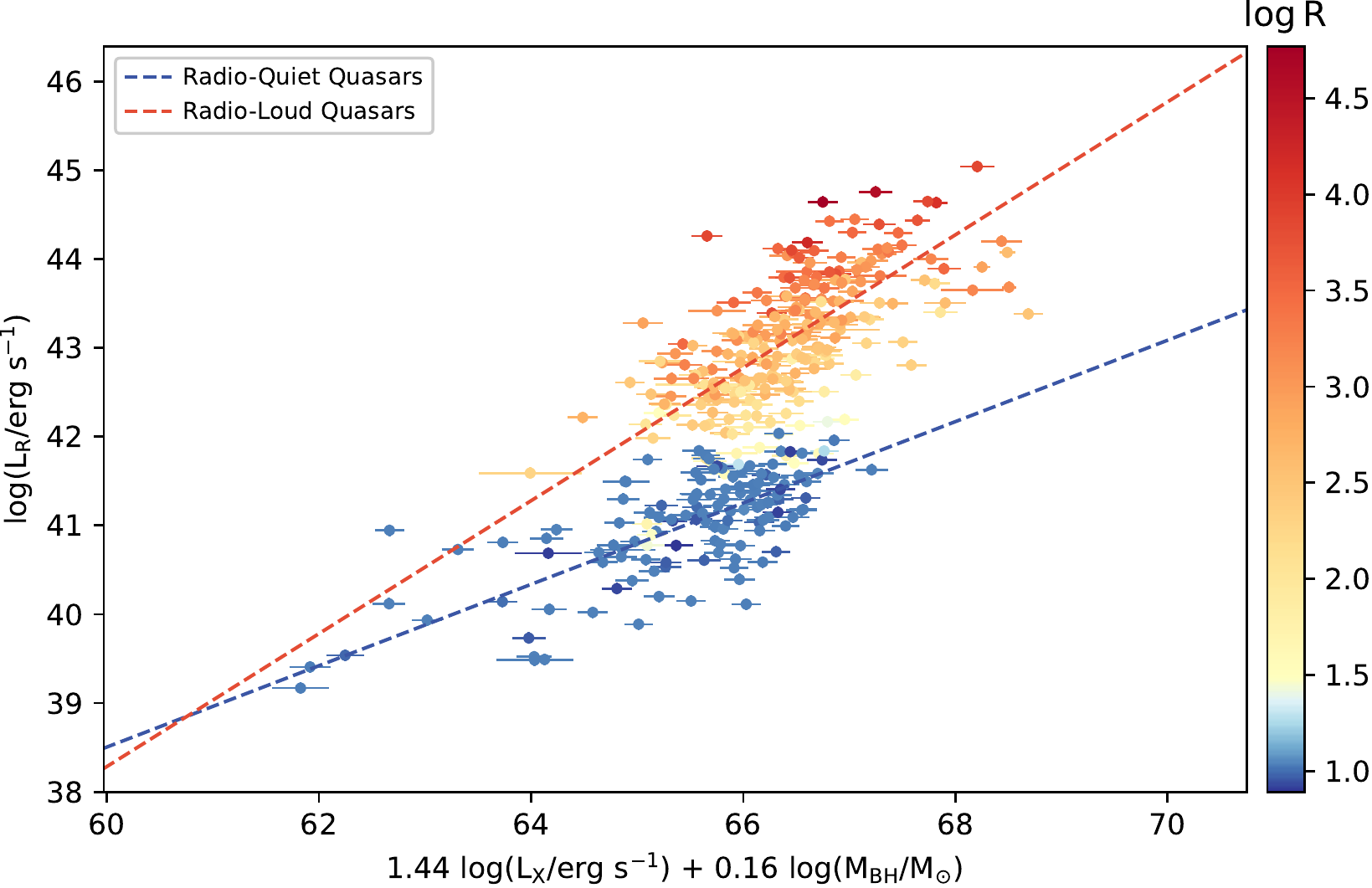}
    \caption{An edge-on view of the fundamental planes of black hole activity, where the sample is separated into radio-loud and radio-quiet quasars. The x-axis is defined in accordance with the best-fit parameters to the total quasar dataset. Best-fit lines for the radio-quiet (blue) and radio-loud (red) are shown. The data demonstrates a clear dichotomy between the fundamental plane best-fit for the two populations.}
    \label{fig:fp_r}
\end{figure*}

For our analysis, we utilized our accumulated samples to construct a fundamental plane of black hole activity. Performing the ODR fit on our radio-loud and radio-quiet sources, we found a fundamental plane best-fit of $\log{L_{R}} = (1.12 \pm 0.06) \log{L_{X}} + (-0.20  \pm 0.07) \log{M} - (5.64 \pm 2.99)$ for our radio-loud sample and $\log{L_{R}} = (0.48 \pm 0.06) \log{L_{X}} + (0.50  \pm 0.08) \log{M} + (15.26 \pm 2.66)$ for our radio-quiet sample. Figure~\ref{fig:fp_r} shows the result of fitting across the two samples, and Table~\ref{table:fundplane} displays the best-fit coefficients. For completeness, a best-fit was obtained for the entire dataset of radio-loud and radio-quiet quasars, which is also provided in Table~\ref{table:fundplane}. The residual variance $\sigma^2_{\rm res}$ and the adjusted coefficient of determination\footnote{We define the coefficient of determination as $R^2 = \frac{r_{xz}^2 + r_{yz}^2 - 2 r_{xy} r_{xz} r_{yz}}{1 - r_{xy}^2}$, where $r_{xy}$, $r_{xz}$, and $r_{yz}$ are the correlation coefficients. Here $x$ and $y$ are the independent variables of the fit, while $z$ is the dependent variable. To account for the number of model parameters in the fit, we calculate the adjusted coefficient of determination \mbox{$R_{\rm adj}^2 = 1 - \frac{(1-R^2)(N-1)}{N-p-1}$}, where $N$ is the total sample size and $p$ is the number of independent variables.} $R_{\rm adj}^2$ were measured for each best-fit model, and the results are shown in Table~\ref{table:fundplane}. A positive correlation is observed for each best-fit, while the fit statistics demonstrate a preference towards utilizing separate models for the radio-loud and radio-quiet samples. 

We compared our best-fit parameters of each sample to the result from \cite{Merloni2003}. The measured scatters of the fit in the log$(L_R)$ parameter relative for radio-loud and radio-quiet quasar samples are $\sigma_R = 0.62$ and $\sigma_{R}=0.39$, which is at least $68\%$ less than the scatter found by \cite{Merloni2003} and at least $75\%$ less than the scatter found by \cite{Gultekin2009}, though this difference may be attributable to our larger adopted uncertainties than those from similar studies. Overall, we found that the radio-quiet and \cite{Merloni2003} best-fit parameters agree with one another within $1\sigma$ despite different data selection criteria (e.g., redshift and mass). This agreement is unsurprising given that the sample from \cite{Merloni2003} was primarily comprised of radio-quiet sources. In comparison, the radio-loud best-fit parameters each diverge from the \cite{Merloni2003} result by $>3\sigma$. Our results suggest that the dichotomy between radio-loud and radio-quiet quasars is also found within the fundamental plane, hence, two fundamental planes are better descriptors of quasar populations. In \S~\ref{sec:discussion}, we discuss the physical impacts of trends observed within the fundamental planes.

%%%%%%%%%%%%%%%%%%%%%%%%%%%%%%%%%%%%%%
\section{Discussion}
\label{sec:discussion}

\begin{figure*}
    \centering
    \includegraphics[width=0.492\linewidth]{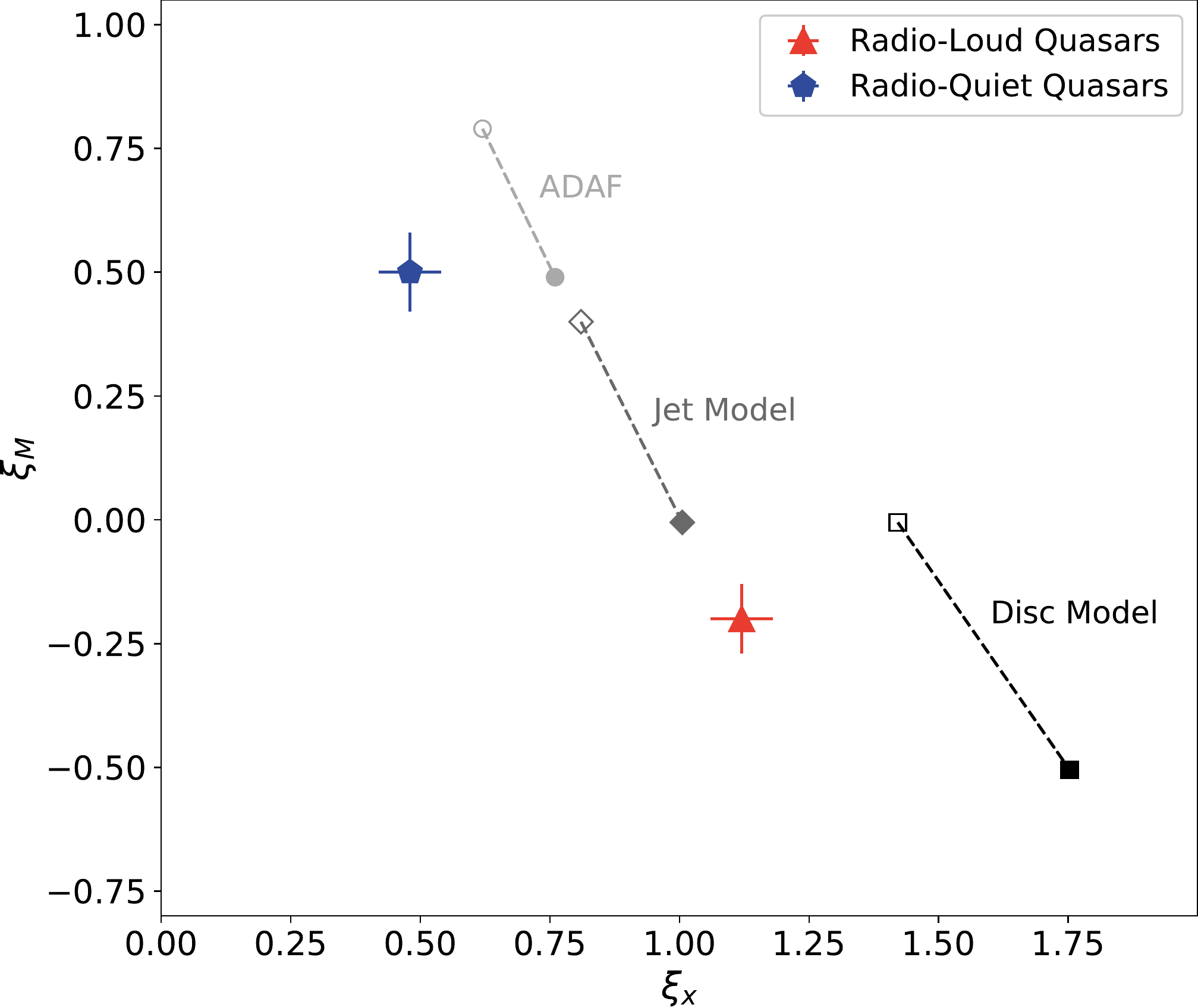}
    \hspace{0.5em}
    \includegraphics[width=0.492\linewidth]{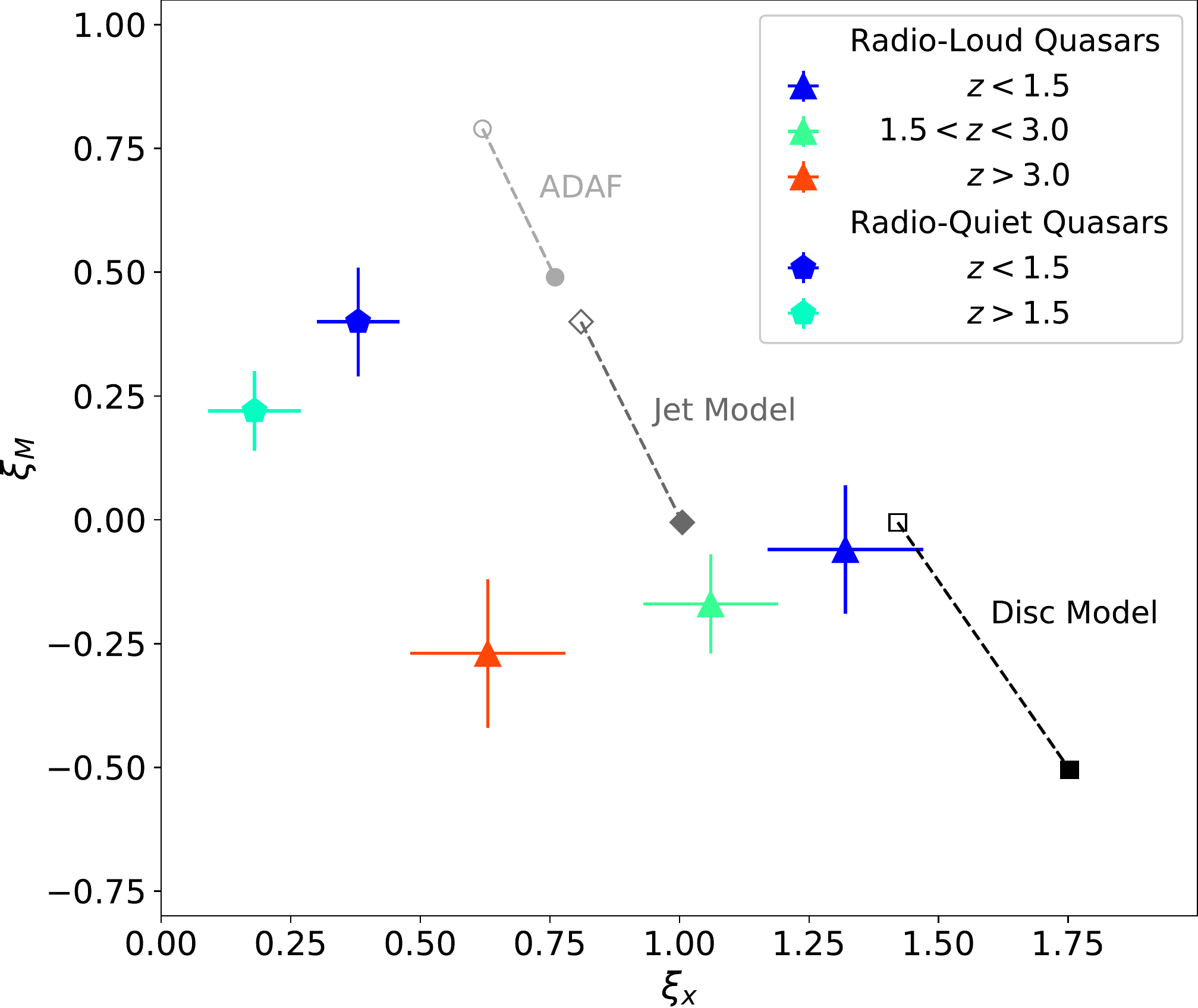}
    \vspace{-1.5em}
    \caption{A comparison of the correlation coefficients $\xi_{\rm X}$ and $\xi_{\rm M}$ for the fundamental plane best-fits of the radio-loud and radio-quiet samples (left) as well as for redshift subsamples (right). Each figure also shows the theoretically predicted correlation coefficients from \citet{Merloni2003}, where circles, diamonds, and squares correspond to Advection Dominated Accretion Flows (ADAF), jet, and disc models respectively.  Empty symbols are for a radio spectral index $\alpha_{\rm R} = 0$ and filled symbols for $\alpha_{\rm R} = 0.5$. The dotted lines represent different possible correlation coefficients due to variations in radio emission.}
    \label{fig:emission}
\end{figure*}

%%%%%%%%%%%%%%%%%%%%%%%%%%%%%%%%%%%%%%
\subsection{Radio-Loud Dependence on Fundamental Plane}
\label{subsec:radioloudness}

The fundamental plane comparisons performed in \S\,\ref{sec:results} reveal several key properties within our quasar samples. First, the fundamental plane from \cite{Merloni2003} with their predominantly radio-quiet heterogeneous sample is consistent with our more homogeneous radio-quiet quasar sample. Second, the \cite{Merloni2003} fundamental plane underpredicts the radio luminosity of the radio-loud quasar sample, resulting in a divergence between the two best-fits over our examined physical parameter space. Radio loudness has previously been shown to impact accretion rates \citep{2007ApJ...658..815S,2012ApJ...759...30B}, so it follows that it should also impact the fundamental plane of black hole activity. Furthermore, studies of radio loudness could help constrain jet formation, acceleration, and collimation \citep{2007ApJ...658..815S}. Motivated by our observation and its physical significance, we explored the fundamental plane within the context of radio loudness. 

The fundamental plane best-fit parameters for the two radio loudness samples are shown in Table~\ref{table:fundplane}, and Figure~\ref{fig:fp_r} displays our results. Comparing the two best-fits, we find that their difference exceeds the measured scatter $\sigma_R$ of 0.62 and 0.39 for the radio-loud and radio-quiet fits, respectively. Thus, we observe a dichotomy between the black hole activity of radio-loud and radio-quiet sources. The observed separation between the radio-loud and radio-quiet samples is consistent with previous studies that have demonstrated a bimodal distribution of radio loudness amongst the quasar population \citep[e.g.,][]{1989AJ.....98.1195K, 1990MNRAS.244..207M,Ivezic2004,2007ApJ...654...99W, 2008MNRAS.387..856Z,2012ApJ...759...30B}. 

The measured correlation coefficients from our fundamental plane best-fits provide insight into the different emission mechanisms within the quasar samples, where the ratio of these observables can correspond to predicted values from different emission models \citep{Merloni2003,2012MNRAS.419..267P}. We therefore plotted the correlation coefficients from our radio-loud and radio-quiet best-fits, which is shown in Figure~\ref{fig:emission},\,left. Joint errors for the $\xi_x$--$\xi_M$ parameters of each best-fit function were calculated from their respective mean covariance matrix, and found to be consistent with their marginalized uncertainties for both the radio-loud and radio-quiet best-fits. The resulting figure further illustrates the dichotomy between our radio-loud and radio-quiet quasar samples as the two best-fits are separated by $>$\,$3\sigma$ for both $\xi_{\rm X}$ and $\xi_{\rm M}$  parameters. Furthermore, these results indicate that the dominant emission mechanism differs between the two samples.

To assess the probable emission mechanism for our quasar samples, we compared our correlation coefficients to those predicted by different emission models from \cite{Merloni2003}. The predicted coefficients are also overlaid on Figure~\ref{fig:emission}. We stress that the plotted model parameters represent a small subsample of the possible quasar emission models, so any emission mechanism predictions and subsequent comparisons should be regarded as approximate. Based on the different model parameterizations from \cite{Merloni2003} that were tested, we found that the radio-quiet sample, on average, favors the advection dominated accretion flows (ADAF) instead of the standard jet or disc models, which agrees with previous fundamental plane studies for radio-quiet black holes \citep[e.g.,][]{Merloni2003,Gultekin2009,2012MNRAS.419..267P}. The distribution of $L_{X}/L_{\rm Edd}$ values for the radio-quiet sample, shown in Figure~\ref{fig:histogramsources}, is also broadly consistent with radiatively inefficient emission based on the model results of \cite{Merloni2003}. In comparison, the radio-loud quasar sample is, on average, more consistent with the standard jet emission model, or a disc model, or a combination of the two. Our finding agrees with previous quasar studies where jet contributions are invoked to explain the radio-loud phenomenon \citep{Worrall1987,1989AJ.....98.1195K,Urry1995}.

Overall, our results demonstrate that two fundamental planes are necessary to accurately reflect the different physical mechanisms that govern radio-loud and radio-quiet quasars. These two separate models will help better constrain and predict physical properties of quasar populations such as the nature of jets, accretion processes, and their local environments.

\begin{figure*}
    \centering
    \includegraphics[width=0.492\linewidth]{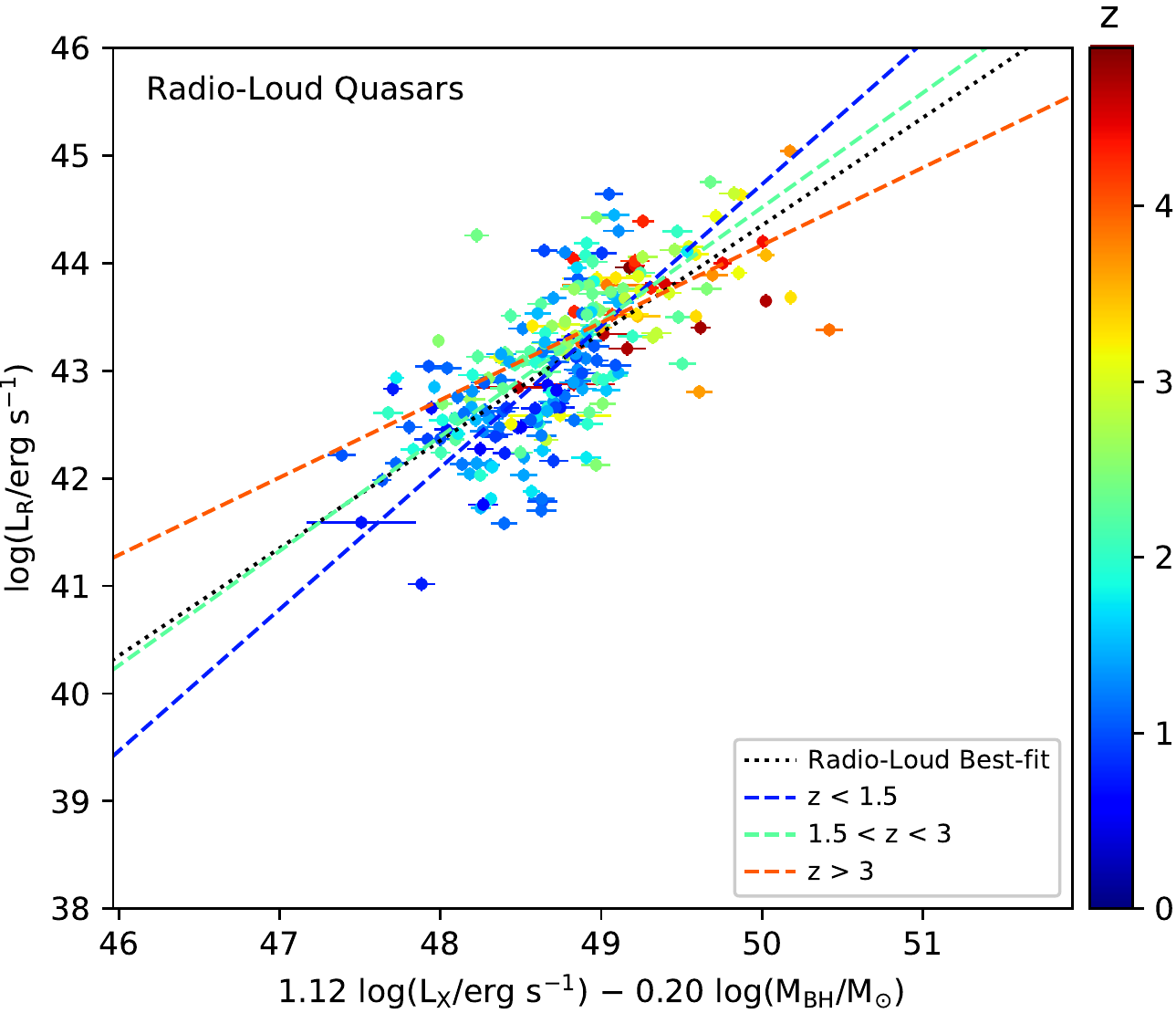}
    \hspace{0.2em}
    \includegraphics[width=0.492\linewidth]{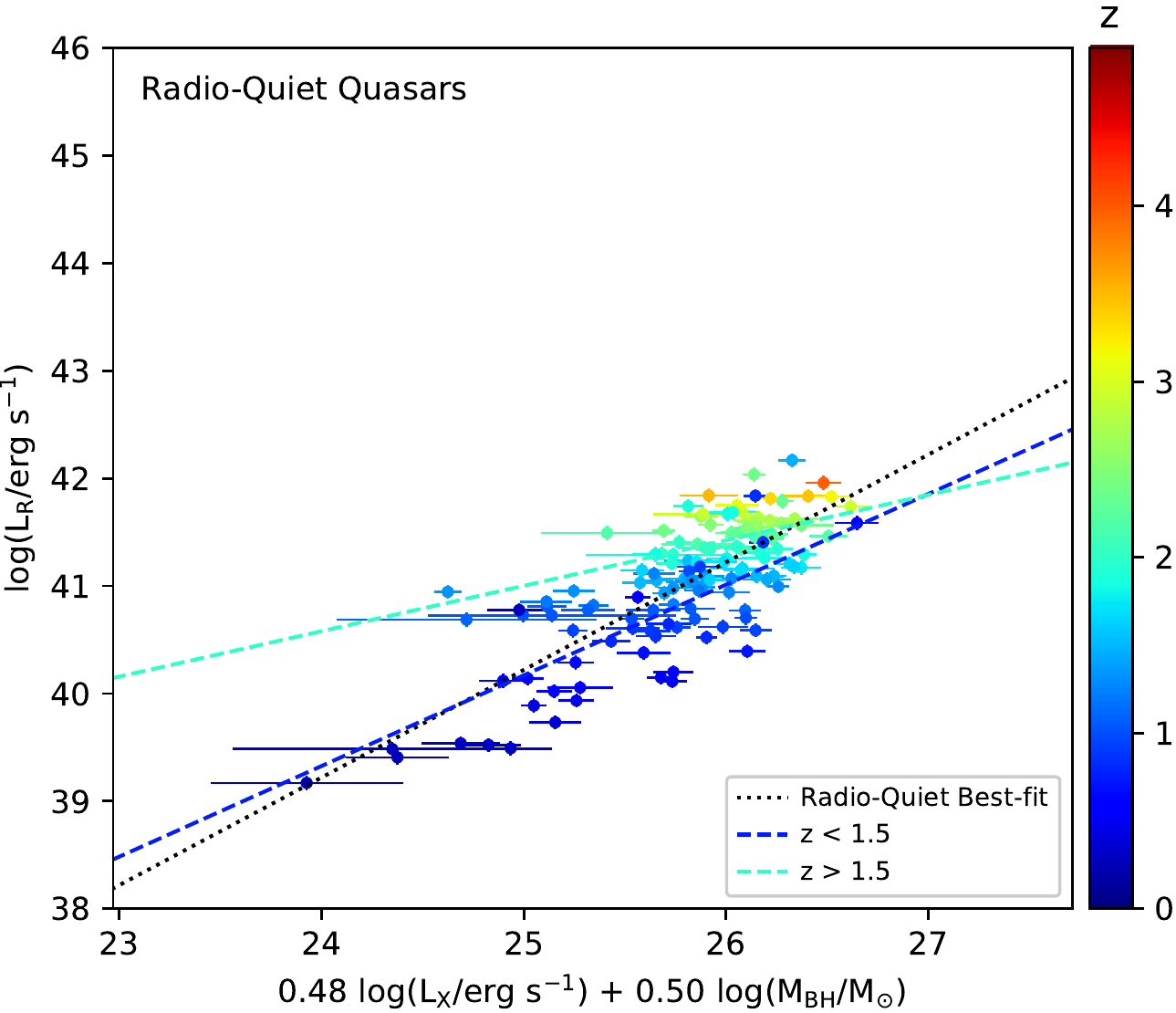}
    \caption{An edge-on view of the fundamental planes of black hole activity across different redshift bins for the radio-loud quasars (left) and radio-quiet quasars (right). The best-fit lines for each redshift bin (colored) and the total sample (black) are plotted. The fundamental plane best-fit parameter are found to be broadly consistent across redshifts for both the radio-loud and radio-quiet samples.}
    \label{fig:fp_z}
\end{figure*}

%%%%%%%%%%%%%%%%%%%%%%%%%%%%%%%%%%%%%%
\subsection{Redshift Dependence on Fundamental Plane}
\label{subsec:redshiftdepedence}

Previous works, due to observational constraints, focused on the fundamental plane within the context of low-redshift sources $z<0.5$ \cite[e.g.,][]{Merloni2003,Falcke2004,2006A&A...456..439K,Gultekin2009}. As a result, the fundamental plane at high-redshifts is not well-understood. Furthermore, whether the fundamental plane model is constant over the age of the Universe is not well-constrained. We therefore investigated the fundamental plane model across varying redshift. 

Due to the observed differences between the population of our radio-loud and radio-quiet sample, we analyzed redshift dependencies of each sample separately. We binned the radio-loud sources across three redshift ranges ($z<1.5$, $1.5<z<3.0$, and $z>3$), where each redshift bin was selected to ensure similar sample statistics for our analysis. In contrast, we binned the radio-quiet sources across two redshift ranges ($z<1.5$, $z>1.5$), due to its smaller sample size. The best-fit results from these additional tests are shown in Table~\ref{table:fundplane}. We note that the measured scatter $\sigma_R$ demonstrates consistency between the different models over the parameter space investigated with our sample (Table~\ref{table:fundplane}). Figure~\ref{fig:fp_z},\,left displays the fundamental plane best-fit for the radio-loud quasars, while Figure~\ref{fig:fp_z},\,right displays the best-fit for the radio-quiet quasars.

From our analysis, the radio-quiet best-fits demonstrate broad consistency with one another across the examined redshift range, where the correlation coefficients are in agreement to within 2$\sigma$ in all scenarios.  Additionally, Figure~\ref{fig:emission},\,right illustrates the consistency between the two radio-quiet subsamples while also indicating that each subsample is most consistent with ADAF from the emission models tested. Our result agrees with optical--X-ray studies of quasars that have shown constant trends in power output and accretion rates within radio-quiet quasars up to redshifts $z\sim6$ \citep{Just2007,Nanni2017,Vito2019b}. Ultimately, our current data which spans a redshift range of $0.1 < z < 5.0$ suggests that the radio-quiet fundamental plane model is independent of redshift. Hence, low-redshift fundamental plane models from previous works \cite[e.g.,][]{Merloni2003,Falcke2004,2006A&A...456..439K,Gultekin2009, 2012MNRAS.419..267P} may be valid for radio-quiet black holes extending all the way to $9\%$ of the present age of the Universe.

In contrast to the radio-quiet sample, examination of the radio-loud quasar subsamples indicates a decreasing X-ray luminosity dependence for increasing redshift at a $>$\,$3\sigma$ significance. This result is also illustrated in Figure~\ref{fig:emission},\,right, where the radio-loud samples clearly demonstrate a decreasing X-ray luminosity dependence versus redshift while mass dependence remains constant to within $1\sigma$. As noted in \S~\ref{sec:dataselection}, the radio luminosity measurements of the high-redshift radio-loud quasar cores may include contributions from extended features due to instrument resolution limits, which would elevate $L_{R}$ with increasing redshift. Comparatively, extended emission, on average, accounts for $<2\%$ the total X-ray luminosity of a quasar \citep[e.g.,][]{Marshall2018,Worrall2020,Snios2021}, which is below other sources of measurement error. If extended emission contamination is present in our radio-loud sample, the $L_{R}$ values of the high-redshift quasars should be regarded as upper limits on the true AGN core luminosity. Reducing $L_R$ to account for such effects would also reduce the measured slope of the fundamental plane for sources at $z>3$, increasing the statistical significance of the inverse relationship between X-ray luminosity and redshift for the fundamental plane. Thus, the presence of contamination would serve to enhance the observed redshift trend for the fundamental plane of the radio-loud quasar population.

Comparing the samples to different emission models from \cite{Merloni2003}, we found that the low-redshift radio-loud sample agrees well with the disc model. However, the high-redshift sample is more consistent with a jet model, indicating a possible evolution of the primary emission mechanism for radio-loud quasars. Such an effect would be consistent with the recent theory that inverse Compton upscattering of the cosmic microwave background radiation (IC/CMB) will increase the observed jet intensity for quasars at $z\gtrsim3$ as the energy density of the CMB exceeds the magnetic energy density for the quasar jets \citep{Schwartz2020,Hodges-Kluck2021}.

We stress that the observed redshift dependence in the radio-loud sample is indicative given the limited sample size and coarse redshift binning utilized in our analysis, and a greater sample size across all redshifts is necessary to accurately constrain the existence of a redshift dependence on the primary emission mechanism. Nonetheless, our analysis does suggest a redshift dependence for the fundamental plane of radio-loud quasars.

%%%%%%%%%%%%%%%%%%%%%%%%%%%%%%%%%%%%%%
\subsection{Using the Fundamental Plane as a Mass Estimator} 
\label{subsec:discussionmass}

\begin{table*}
    \caption{ Best-Fit Parameters of Black Hole Mass Fundamental Plane Analysis}
    \label{table:fundplanemass}
    \begin{tabular}{lcrrrrrr}
        \hline
        \hline
         & Number of Sources & \multicolumn{1}{c}{$\xi_{MX}$} & \multicolumn{1}{c}{$\xi_{MR}$} & \multicolumn{1}{c}{$b_{M}$} &  \multicolumn1c{$\sigma^2_{\rm res}$} & \multicolumn1c{$R_{\rm adj}^2$} & \multicolumn{1}{c}{$\sigma_{M}$}\\
        \hline
        Radio-Loud Quasars & 225 & $0.63 \pm 0.05$ & $-0.21 \pm 0.04$ & $-10.47 \pm 2.09$ & $15.56$ & $0.1474$ & $0.43$\\
        Radio-Quiet Quasars & 128 & $0.08 \pm 0.07$ & $0.41 \pm 0.06 $ & $-11.42 \pm 3.61$ & $32.00$ & $0.3281$ & $0.45$ \\
        \hline
    \end{tabular}
    \\\raggedright{{\sc note}: The reported parameters for the fundamental plane best-fits correspond to the X-ray luminosity slope $\xi_{MX}$, the radio luminosity slope $\xi_{MR}$, the y-intercept $b_{M}$, the best-fit residual variance $\sigma^2_{\rm res}$, the adjusted R-squared of the best-fit model $R_{\rm adj}^2$, and the scatter of the log$(L_R)$ parameter relative to the predicted value from the fundamental plane models $\sigma_{M}$.}
\end{table*}

\begin{figure*}
    \centering
    \includegraphics[width=0.49\linewidth]{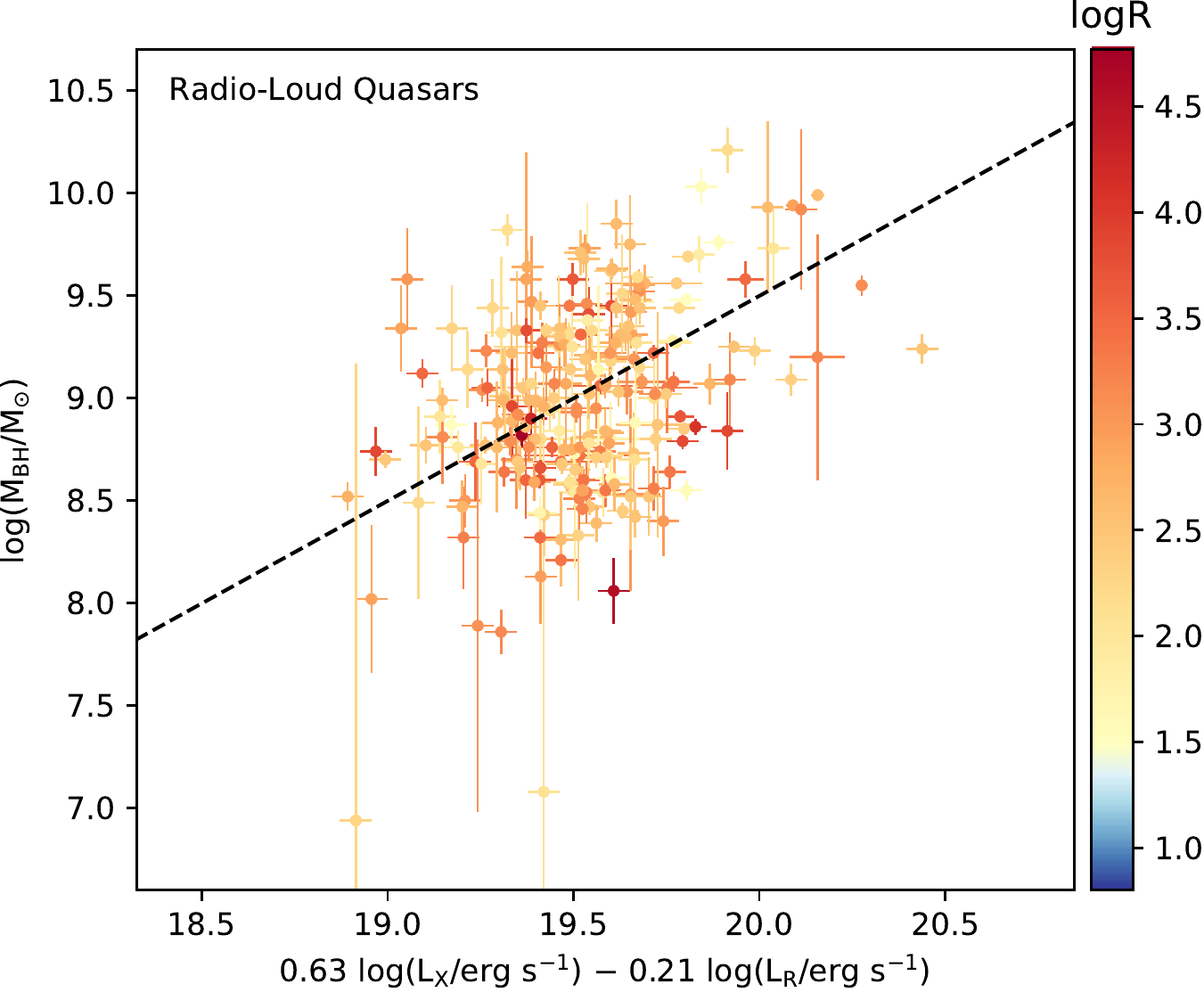}
    \hspace{0.5em}
    \includegraphics[width=0.49\linewidth]{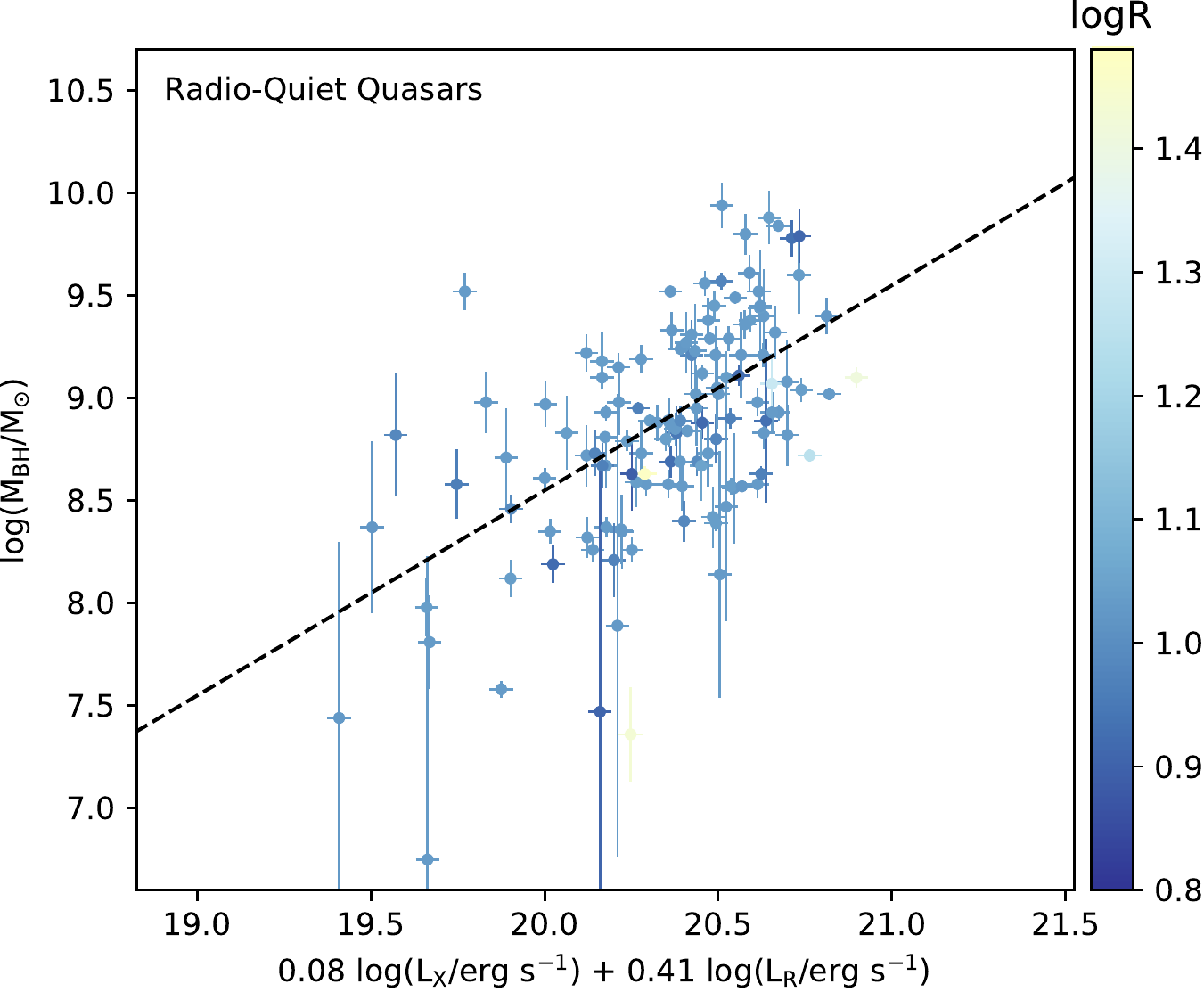}
    \caption{An edge-on view of the fundamental planes of black hole activity where $\log{M_{BH}}$ is the predicted variable, where the sample is separated into radio-loud (left) and radio-quiet quasars (right). Best-fit lines for the radio-quiet and radio-loud are shown. The derived mass measurements are found to be consistent with spectroscopic mass estimation techniques for black holes, where the accuracy of the fundamental plane mass estimates is comparable to those obtained from spectroscopic mass measurements of high-redshift quasars.}
    \label{fig:fp_mass}
\end{figure*}

Historically, fundamental plane studies have focused on investigating the accretion properties of black holes across broad mass and luminosity ranges. However, the fundamental plane may also be used as an estimator of black hole masses due to the scale-invariant nature of the model \citep[e,g.,][]{Merloni2003,2019ApJ...871...80G}. Motivated by this fact, we examined the fundamental plane in the context of mass prediction for our quasar sample. 

Using the ODR fitting method described in \S\,\ref{sec:results}, we defined our model as the expression
\begin{equation}
    \log{M} = \xi_{MX} \log{L_{X}} + \xi_{MR} \log{L_{R}} + b_{M}.
    \label{eqn:fundplanemass}
\end{equation}
Due to the observed dichotomy between radio-loud and radio-quiet quasars described in \S\,\ref{subsec:radioloudness}, we fit the model to each sub-sample separately. The results for all our fits are shown in Table\,\ref{table:fundplanemass}, and Figure \ref{fig:fp_mass} displays the best-fit results of our analysis. The results indicate a dichotomy between the radio-loud and radio-quiet sample when fitting the fundamental plane defined in Equation~\ref{eqn:fundplanemass}, where the best-fit slope parameters $\xi_{MX}$ and $\xi_{MR}$ differed by $>3\sigma$ between the two best-fits. The measured scatter of the log$(M)$ parameter between the measured and predicted values $\sigma_M$ are determined to be 0.43 and 0.45 for the radio-loud and radio-quiet samples, respectively. The scatters are consistent between the two datasets, and both samples cover a comparable physical parameter range. Altogether, we find that the fundamental plane model for mass measurements is dependent on radio loudness. 

To assess the accuracy of the fundamental plane as a black hole mass predictor, we compared the predicted masses determined by our fundamental plane fits with the spectroscopically measured values from the SDSS-DR14 emission line analysis compiled by \cite{2020ApJS..249...17R}. We therefore calculated the root mean square error\footnote{We define the root-mean square error as $[\sum_{i}^{N}(O_{i}-P_{i})^2/N]^{1/2}$, where $O$ is the measured value from the fundamental mass plane, $P$ is the mass value from \cite{2020ApJS..249...17R}, and $N$ is the sample size.} (RMSE), also known as the standard deviation of the residuals, between the predicted black hole mass from the fundamental plane model and the spectrum-measured mass from \cite{2020ApJS..249...17R}. We found that the radio-loud and radio-quiet quasars have an RMSE of $0.43$\,log($M_{\odot})$ (5.0\%) and $0.45$\,log$(M_{\odot})$ (5.5\%), respectively. Our method of black hole mass estimation is broadly consistent with previous studies that utilized the fundamental plane as a mass estimator \citep[e.g.,][]{2019ApJ...871...80G}, despite our study focusing on high-mass quasars ranging between $10^{6}$\,--$10^{10} M_\odot$ while previous works prioritized low-mass, low-redshift sources.  

The accuracy of black hole mass estimates from spectroscopic modeling is known to worsen with increasing redshift due to a reduction in the available emission lines in the optical band as well as a greater reliance on low signal-to-noise emission lines, such as Mg\,II and C\,IV \citep{Shen2013}. We theorize that the mass estimates from the fundamental plane would not suffer from a similar degradation in accuracy with increasing redshift given the comparable radio and X-ray flux errors across our sample. To test this theory, we divided our total quasar sample into low-redshift ($z<3$) and high-redshift ($z>3$) subsamples. For the fundamental plane mass measurements, we found an average RMSE of 5.1\% for quasars at $z<3$ and 5.4\% for quasars at $z>3$. In contrast, the spectroscopic mass measurements taken from \cite{2020ApJS..249...17R} have an average error percentage of 1.6\% for quasars at $z<3$ and 2.5\% for quasars at $z>3$. 

Given our selection of conservative errors for the radio and X-ray fluxes (\S~\ref{sec:results}), the precision of the fundamental plane method could improve if more accurate uncertainties are utilized. Nonetheless, our results demonstrate that the accuracy of the fundamental plane mass model is predominantly redshift invariant, while the accuracy of spectrum-measured masses worsens with redshift. Thus, our derived fundamental plane for black hole mass may prove useful in studying high-redshift quasars as well as situations where more standard mass estimation techniques, such as spectral modeling, are not possible. 

%%%%%%%%%%%%%%%%%%%%%%%%%%%%%%%%%%%%%%
\section{Conclusion}
\label{sec:conclusion}
We examined the fundamental plane of black hole activity for correlations with redshift and radio loudness in both radio-loud and radio-quiet quasar populations. Utilizing archival observations, we compiled quasar data that varied across radio loudness, mass, and redshift to obtain a sample of 353 sources. We constructed a fundamental plane of black hole activity, and our best-fits are $\log{L_{R}} = (1.12 \pm 0.06) \log{L_{X}} - (0.20  \pm 0.07) \log{M} -(5.64 \pm 2.99)$ for our radio-loud sample and $\log{L_{R}} = (0.48 \pm 0.06) \log{L_{X}} + (0.50  \pm 0.08) \log{M} + (15.26 \pm 2.66)$ for our radio-quiet sample.

We examined the impact of radio loudness on the fundamental plane best-fit for within our quasar sample. Our results demonstrate that two different fundamental planes are required to accurately model the dichotomy of radio-loud and radio-quiet quasar populations. Thus, we concluded that the radio-loud and radio-quiet quasar populations are governed by different fundamental plane models. Comparing the best-fit correlation coefficients to values predicted from different emission models, we found that a possible emission mechanism for the radio-quiet sample is advection dominated accretion flows while the radio-loud sample is likely a combination of jet and disc emission.

We also explored the dependencies of redshift on the fundamental plane model for our samples of radio-loud and radio-quiet quasars. Our derived best-fit results show no redshift dependence for radio-quiet quasars $z \leq 5$, where all correlation coefficients agreed within $2\sigma$. In contrast, the radio-loud quasars exhibit a decreasing X-ray luminosity dependence for increasing redshift at a $>3\sigma$ level, which we theorize may be due to the primary emission mechanism of radio-loud quasars evolving over redshift. However, a greater sample size is necessary to conclusively verify the existence of, or lack thereof, an evolving emission mechanism for the radio-loud quasar population.

Leveraging our fundamental plane results, we derived a best-fit expression for black hole mass estimation that used only the radio and the X-ray luminosities. We found a best-fit of $\log{M} = (0.63\pm0.05) \log{L_{X}} - (0.21 \pm 0.04) \log{L_{R}} - (10.47 \pm 2.09)$ for the radio-loud quasars and $\log{M} = (0.08\pm0.07) \log{L_{X}} - (0.41 \pm 0.06) \log{L_{R}} - (11.42 \pm 3.61)$ for the radio-quiet quasars. The accuracy of our fundamental plane mass estimation method was assessed for each sample, and we found an average $4.0\%$ difference between it and standard estimates from optical spectrum emission lines. Thus, this method may prove useful in studying the high-redshift quasar population as well as situations where high signal-to-noise emission lines utilized for spectral modeling are not present at standard observing wavelengths.

Follow-up studies with more robust statistics may be used to better constrain model uncertainties. Furthermore, additional sources from future large-scale quasar surveys, such as the VLA and eROSITA all-sky surveys that are currently in progress, will increase the number of available targets and improve the the statistical significance of future fundamental plane studies. 

\section*{Acknowledgements}

L.B. was supported by the SAO REU program, which is funded in part by the National Science Foundation REU and Department of Defense ASSURE programs under NSF Grants no.\ AST 1852268 and 2050813, and by the Smithsonian Institution. B.S., M.S., A.S., and D.A.S. were supported by NASA contract NAS8-03060 ({\it Chandra} X-ray Center). B.S. was also supported in part by CXC grants GO8-19093X and GO0-21101X. 

L.B. also thanks Matthew Ashby and Jonathan McDowell for their support and mentorship throughout the SAO REU program.

\section*{Data Availability}

The data used in this investigation are available with the article in a machine-readable format.

\bibliographystyle{mnras}
\bibliography{bibliography}

% Don't change these lines
\bsp	% typesetting comment
\label{lastpage}
\end{document}